\documentclass[12pt,preprint]{aastex}

\def\ltorder{\mathrel{\raise.3ex\hbox{$<$}\mkern-14mu
             \lower0.6ex\hbox{$\sim$}}}
\def\lsim{\lower.5ex\hbox{$\; \buildrel < \over \sim \;$}}

\begin{document}


\title{Polarization of the Crab pulsar and nebula as observed by the Integral/IBIS telescope}

 \author{M. Forot\altaffilmark{1,3}, P. Laurent\altaffilmark{1,2}, I. A. Grenier\altaffilmark{1,3}, C. Gouiff\`es\altaffilmark{1,3}, F. Lebrun\altaffilmark{1,2}}

\altaffiltext{1}{CEA,IRFU,Service d'Astrophysique, 91191 Gif sur Yvette, France;plaurent@cea.fr}
\altaffiltext{2}{APC, 10 rue Alice Domont et Leonie Duquet, 75205 Paris Cedex 13, France}
\altaffiltext{3}{AIM, CEA/CNRS/Universit\'e Paris Diderot, SAp, Saclay, 91191 Gif-sur-Yvette, France}

\begin{abstract}
Neutron stars generate powerful winds of relativistic particles that form bright synchrotron nebulae around them. Polarimetry provides a unique insight into the geometry and magnetic configuration of the wind, but high-energy measurements have failed until recently. The Integral-IBIS telescope has been used in its Compton mode to search for linearly polarized emission for energies above 200 keV from the Crab nebula. The asymmetries in the instrument response are small and we obtain evidence for a strongly polarized signal at an angle parallel to the pulsar rotation axis. This result confirms the detection recently reported by Dean et al. (2008), and extends the polarization measure for all the pulsar's phases. The hard X-ray/soft $\gamma$-ray observations therefore probe the inner jets or equatorial flow of the wind. The achieved sensitivity opens a new window for polarimetric studies at energies above 200 keV.
\end{abstract}

\section{Introduction}
The Crab nebula is the spectacular remnant of a star that exploded in 1054. The leftover neutron star accelerates particles up to nearly the speed of light and flings them out into the nebula at energies up to $10^{15}$ eV. As they age, the bright synchrotron radiation is seen at all wavelengths and can be used for polarimetry to explore the wind geometry and its magnetic field, in order to constrain where and how such efficient acceleration takes places. Polarization has been measured from the radio to X rays with energy up to 5.2 keV (Weisskopf et al. 1978). Recently, polarization has also been detected in the 100 keV - 1 MeV range  using Integral/SPI data (Dean et al., 2008). The high energy detections prove to be very useful since hard X rays/soft $\gamma$-rays are produced by the highest-energy particles, freshly ejected by the pulsar.

$\gamma$-ray polarimetry has been possible with Compton telescopes since the 1970s. Photons that are Compton scattered between two detectors follow an azimuthal distribution around the source direction that allows to quantify the degree and direction of linear polarization because the photon is preferentially scattered in a plane at right angle to its incident electric vector. It was unsuccessful up to recently because of intrinsic asymmetries in the detector response and of non-uniformities in the large background signals. They induce pseudo polarimetric signals, even from an unpolarized source, that limit the sensitivity to any detection. For instance, attempts at detecting polarization in bright $\gamma$-ray bursts have been met with varying success and need confirmation (McGlynn et al. 2007 and references therein). With its double layer of finely pixellated detectors (Ubertini et al. 2003), the IBIS telescope on board the Integral satellite is well suited for polarimetry studies, between 200 keV and 5 MeV. The background is efficiently subtracted by deconvolving the coded-mask shadowgram on the primary detector. The angular resolution is energy independent and the two detector planes are close enough to detect events scattered at large angles with a rather uniform response in azimuth.

We first describe the polarimetry method we have developed, and check the level of asymmetries in the instrument response using different unpolarized sources. We then present evidence for the detection of polarization from the Crab nebula, at energies above 200 keV. Finally, we discuss the possible origin of the signal.

\section{Polarimetry with the Integral/IBIS Compton mode}

Photons entering IBIS are Compton scattered in the first detector plane, ISGRI (Lebrun et al., 2003), at a polar angle $\theta$ from their incident direction and at an azimuth $\psi$ from their incident electric vector. They are then absorbed in the second detector, PiCsIT (Labanti et al., 2003). The azimuthal profile $N(\psi)$, in Compton counts recorded per azimuth bin, follows:
\begin{equation}
N(\psi) = S [1+ a_0 cos(2\psi-2\psi_0)]
\end{equation}
\noindent
 for a source polarized at an angle PA = $\psi_0 - \pi/2 + n\pi$ and with a polarization fraction PF = $a_{0}/a_{100}$. The $a_{100}$ amplitude is expected for a 100\% polarized source. Unfortunately, the IBIS polarimetric capacities have not been calibrated on ground, due to the tight planning of space missions. We have then evaluated $a_{100}$ to be $0.30 \pm 0.02$ for a Crab-like $E^{-2.2}$ spectrum between 200 and 800 keV, using GEANT3 Monte-Carlo simulations of IBIS and its detailed mass model (Laurent et al., 2003), using the GLEPS package\footnote{GLEPS is a package for handling polarization in Geant 3 developed by Dr. Mark McConnell at University of New Hampshire, USA} for polarization. This value agrees with the early estimate of 0.3 obtained in the 200-500 keV band by Lei et al. (1997).

Events recorded in ISGRI and PiCsIT within the same time window of 3.8 $\mu$s are tagged as 'Compton' events, but do not all result from Compton scattering. Chance coincidences can occur between ISGRI and PiCsIT events independently coming from the source, the sky, or the instrumental background. These coincidences are generally called spurious events. Most of the 'Compton'-tagged events are due to background events that will be removed by the shadowgram deconvolution; 5\% are due to a small fraction of spurious coincidences that must be removed with high accuracy because they induce a false source detection in the sky image; (Indeed, among the spurious events, there are some for which the ISGRI or PiCsIT events are really coming from the source : for instance, an ISGRI low energy Crab event could be associated by chance with a PiCsIT background event; these events coming from the source are then not removed by the deconvolution process.) 2\% come from true Compton events from the source.

 To measure $N(\psi)$, we first optimize the true-Compton to spurious ratio by applying angular cuts on the difference between the true source direction and the reconstructed one (from the events geometry and energies, assuming Compton scattering). The remaining spurious counts described above are estimated by randomizing in time all the ISGRI and PiCsIT events recorded during the observation (Forot et al. 2007). After subtraction of these spurious counts, the ISGRI shadowgram is deconvolved to remove the background and get source counts. The whole process is applied for events in regularly spaced bins in azimuth to derive $N(\psi)$. The errors on $N(\psi)$ are dominated by statistic fluctuations in our background dominated observations. To improve the polarimetric sensitivity, we keep only fully-coded observations, at off-axis angles $< 5^\circ$.

Confidence intervals on $a_0$ and $\psi_0$ are not given by the $N(\psi)$ fit to the data since the variables are not independent. They have been derived from the probability density distribution of measuring a and $\psi$ from $N_{pt}$ independent data points in $N(\psi)$ over a $\pi$ period, based on Gaussian distributions for the orthogonal Stokes components (Weisskopf 2006, Vaillancourt 2006, Vinokur 1965):

\begin{equation}
dP(a,\psi) = \frac{N_{pt}~S^2}{\pi~\sigma_S^2} exp[-\frac{N_{pt}~S^2}{2~\sigma_S^2}[a^2+a_0^2-2aa_0cos(2\psi-2\psi_0)]]~a~da~d\psi
\end{equation}
\noindent

 $\sigma_S$ notes the error on the profile mean S. The errors on each a or $\psi$ dimension are obtained by integrating dP(a,$\psi$) over the other dimension.

\section {Calibration and performances of the IBIS/Compton polarimeter}

To avoid a false polarimetric claim, we have looked for residual modulation of instrumental origin in $N(\psi)$ for a series of unpolarized sources, using on-axis and off-axis radioactive calibration sources, empty fields and spurious samples. There is an obvious projection effect for off-axis sources. We have analyzed the data obtained from an unpolarized 392 keV source inclined at $30^\circ$ during the on-ground calibration campaign. It exhibits a clear cosine profile with period $2\pi$ because of the projection effect on the detectors. Correcting analytically this effect, and folding this profile over $\pi$ cancels the observed modulation, while preserving the polarization pattern as it has the same $\pi$ period.

The non-axisymmetric geometry of the detectors (grids, corners, square mask pattern) can also induce a small modulation. Its amplitude has been measured from the data of bright on-axis calibration sources at 392 and 662 keV. Source counts dominate over spurious and background ones. We find modulation fractions $PF = 0.066 \pm 0.013$ and $0.049_{-0.013}^{+ 0.016}$, respectively, indicating that the detectors square geometry induces a significant, but small modulation at an angle near $40^\circ$, primarily because of the sensitivity difference between two adjacent PiCsIT modules. In these studies, the reference frame has been oriented to allow a direct angle comparison with the Crab data below.

In order to check for systematics due, for instance, to the background or the analysis process, we have computed the polarimetric pattern of a "pseudo source" located $1.5^\circ$ away from the Crab, that is out of the Crab point spread function but still in the IBIS field of view. We used the same set of observations and analysis software as the ones used for the Crab polarimetric measurements. As the source is not real, its mean count rate was consistent with zero. We have then measured the level of modulation around zero due to potential statistic plus possible systematic fluctuations in the polarimetric pattern. We got to the conclusion that the fluctuations were consistent with statistics ones, so that the possible systematic fluctuations can be neglected given the statistic level of our observations.

Despite the efficient subtraction of spurious events, a small residual contamination may also modify the shape of $N(\psi)$. We have analyzed a sample of in-flight spurious events recorded between 200 and 800 keV during Crab observations. The profile, displayed in Fig. \ref{data}, shows a small and well defined modulation with PF = $0.15 _{-0.03}^{+0.04}$ at an angle of $175.1° \pm 7.8°$ due to the detectors segmentation. The probability for PF to be greater than 0.15 is $1.2~ 10^{-3}$.

\section{Data observations and discussion}

\begin{figure}
   \plotone{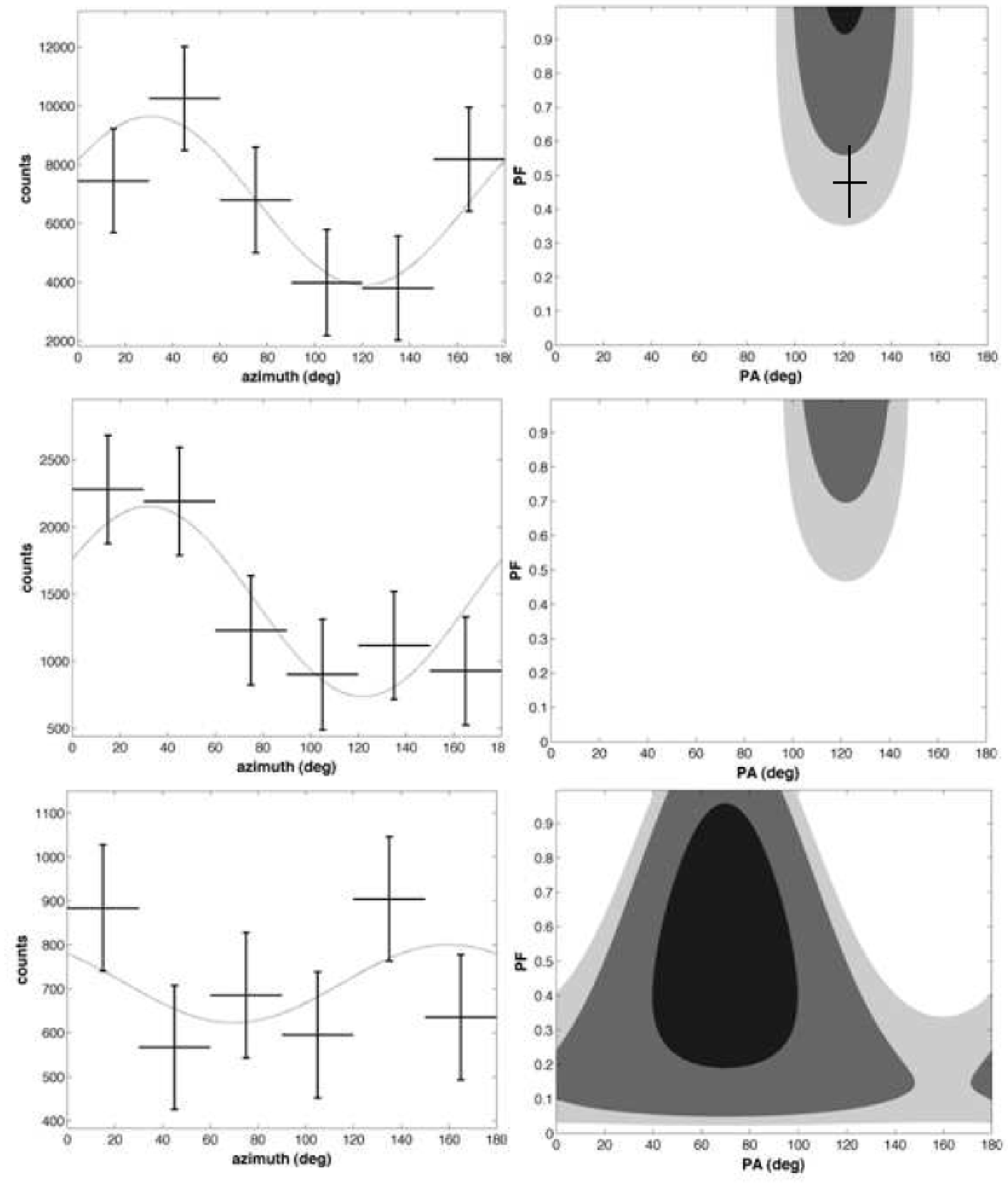}
   \caption{Azimuthal profile, modulation angle, PA, and fraction, PF = $a_0/a_{100}$, measured for the Crab data between 200 and 800 keV, in the off-pulse (top), off-pulse and bridge (middle), and two-peak (bottom) phase intervals. The error bars for the profile are at one sigma. The 68\%, 95\%, and 99\% confidence regions are shaded from dark to light gray. The SPI result (Dean et al., 2008) is indicated in the top figure by a cross.}
   \label{data}
   \end{figure}

The Crab nebula has been repeatedly observed by Integral between 2003 and 2007 and a total of 1.2 Ms of fully-coded observations can be used for polarimetry. The pulsed lightcurve in the 200-800 keV band has been constructed with the Jodrell Bank ephemerides of the pulsar (Lyne et al., 1993). We have considered four phase intervals for polarimetry studies(see table \ref{phases}) : the two main peaks which are dominated by pulsed emission born inside the light cylinder (the pulsed/DC count ratio here is 2) ; the off-pulse interval dominated by nebular emission; and the bridge interval, also dominated by nebular emission. The interval boundaries were taken from (Kuiper et al. 2001) and were not adapted to enhance a possible signal. Table \ref{results}  and Fig. \ref{data}  show the results and illustrate the contrast between the modulation obtained for the nebular emission versus the flat profile of the pulsed emission.

There is no significant indication of polarization in the pulsed peaks. The chance probability of a random fluctuation reaches 33.5 \% and the signal shows no modulation at the 95 \% confidence level over all angles. This behaviour is consistent with the radio and optical data where PF drops below 10 \% as the angle largely flips within each peak (Slowikowska et al. 2006, 2008). Conversely, the chance is low that the modulation seen in the off-pulse emission above 200 keV be of random origin. The P(a,$\psi$) probability density yields a probability of $2.6 ~10^{-3}$ that a random fluctuation produces an amplitude $a_0$ larger than the recorded one. Adding the bridge and off-pulse data strengthens the signal and gives a chance probability of $10^{-3}$ that an unpolarized source produce this modulation. The observed modulation strongly differs from that of the spurious events recorded in the same Crab observations and from the instrumental asymmetries. So, the evidence for a strongly polarized signal holds against both statistical and systematic uncertainties. The polarization of the DC emission appears to be strong enough to still yield a marginal signal in the total Crab emission (see table \ref{results}).

The off-pulse-and-bridge emission is polarized at an angle of $122.0^\circ \pm 7.7^\circ$ which is fully consistent with the north-to-east angle $\psi = 124^\circ\pm 0.1^\circ$ of the pulsar rotation axis projected on the skyplane (Ng \& Romani 2004). The large data sample was taken in different observing modes, none of which was particularly aligned with the pulsar axis in the sky. At the 95 \% confidence level, we find fractions $PF > 72 \%$ and $PF > 88 \%$ in the off-pulse and off-pulse-and-bridge emission, respectively, for any values of the polarization angles. These values, which also do not take into account the incertainties in the $a_{100}$ computation, are both consistent, at the 95 \% level, with an 77 \% polarized signal along the pulsar rotation axis, which is the maximum polarization fraction allowed for synchrotron radiation in a uniform magnetic field and from a power-law distribution of electrons with the spectral index p = $3.454 \pm 0.026$ recorded at these energies (Kuiper et al. 2001).

Our off-pulse measurement confirms the recently observed polarization in hard X-ray using Integral/SPI data (Dean et al., 2008). Indeed, they measure in the off-pulse region, between 100 keV and 1 MeV, an angle of $123.0^\circ \pm 11^\circ$, very close to our own values (see table \ref{results}). However, their polarization fraction $46 \pm 10 \%$ seems marginally consistent with our lower limits. However, our lower limit, as noticed previously, is computed for any value of the polarization angle; the SPI and IBIS results are consistent at the 95 \% level, if we fix the angle at the SPI measured best fit value, as it can be seen on figure \ref{data}.

The angle and large polarization fraction match the observed optical polarization at the same off-pulse phase within 0.01 pc from the pulsar ($0.78 < \phi < 0.84$, PF= 0.33 and PA= $118.9^\circ$ within 1.15", Slowikowska et al. 2006, 2008; PF = $0.47 \pm 0.1$ and PA= $130^\circ$ within 1", Smith et al. 1988). Larger angles and lower fractions have been measured further away from the pulsar. The percentage falls to $8.1 \pm 0.4 \%$ and the orientation rotates to $152^\circ \pm 2^\circ$ in the optical at a distance of 0.02-0.04 pc from the pulsar (Smith et al. 1988, Slowikowska et al. 2006, 2008). The optical polarization of 12.4 \% at $146^\circ$ found at 0.06 pc toward the south-west (Hickson \& van den Bergh 1990) also applies to the wisps region and as far as 0.3-0.5 pc. They correspond to the radio data (Velusamy 1985). At much larger scale, the visible and radio polarisations increase near the edge of the nebula, with electric vectors generally pointing outward. X rays from the whole nebula are $19.2 \pm 1.0$ \% polarized at $156.4^\circ \pm 1.4^\circ$ at 2.6 keV, and $19.5 \pm 2.8$ \% polarized at $152.6^\circ \pm 4.0^\circ$ at 5.2 keV (Weisskopf et al. 1978).

As in X rays, the IBIS measurement encompasses the entire nebula and pulsar. Yet, the nebular spectrum is known to soften with distance from the pulsar. The highest-energy particles flow from the equatorial wind near the termination shock (DelZanna et al. 2006). The synchrotron radiating electrons traced by IBIS have energies as high as 250-500 TeV in the average $16.2 \pm 1.8$ nT nebular field (Aharonian et al. 2004). They live for 0.85-0.43 year and do not travel beyond 0.09 pc downstream of the shock. Their energy and the agreement between the hard X-ray polarization and the optical one from the central 0.01 pc region suggests that the polarized emission seen by IBIS mostly come from the regions well inside the X-ray ring (supposedly the wind termination shock) and its wisps. Any toroidal magnetic configuration (jet, torus, Parker spiral) would exhibit a polarization angle parallel to the pulsar spin axis.

Different sites can spawn unpulsed polarized light at high energy. Within the light cylinder, the slot-gap and extended outer-gap models (Dyks et al. 2004,Takata et al. 2007) predict a small contribution of highly-polarized DC emission for the magnetic obliquity ($60^\circ$) and observer inclination $\xi = 61.3^\circ \pm 0.1^\circ$ of the Crab pulsar (Ng \& Romani 2004), but the predicted angles are at variance with the optical and IBIS data ($20^\circ \leq PA \leq 40^\circ$ for the slot gap, and PA $\approx \xi$ for the outer gap). In the striped wind zone, just outside the light cylinder (Kirk \& Lyubarsky 2001,Petri \& Kirk 2005), particle acceleration and pulsed high-energy emission occur in a thin layer along the equatorial plane where the toroidal magnetic field from the oblique rotator reverses its orientation and dissipates. Faint DC emission outside the broad peaks is possible with $30-40 \%$ polarization aligned with the rotation axis, in agreement with the optical and IBIS data. The polar jets and inner equatorial wind are also sources of polarized radiation. The ratio of pulsar DC emission to wind radiation is unknown. The central optical measurement includes the bright knot discovered along the jet axis, 0.65" south-east of the pulsar (Hester et al. 1995). It has been interpreted as Doppler boosted radiation from the inner part of the arch shock that forms on the upper and lower sides of the equatorial flow, near the pulsar, well inside the torus-like termination shock that forms in the equatorial plane (Komissarov \& Lyubarsky 2004). MHD models predict that polarization is strongest at the pulsar, in the knot, and along the jets and that it should be mostly parallel to the rotation axis as in the optical and IBIS data (DelZanna et al. 2006). So, the off-pulse polarized emission recorded above 200 keV can come from the striped wind, jets, and/or the equatorial wind near the bright knot. Confirming with more statistics the apparent relation between the optical and hard X-ray polarization, and testing the frequency dependence of the polarization pattern, will be crucial to probe the complex structure of the inner wind region.

\section{Conclusions}
The Crab nebula is the brightest persistent source in the sky above 200 keV. The significant, but modest, detection of polarized light from it, in parallel to the SPI detection, opens a new window for high energy astrophysics. The measured polarization angle, parallel to the pulsar rotation axis and similar to the optical measured ones, strongly suggest that the hard X-ray emission is produced well inside the pulsar nebula. No significant polarized emission was found in the pulsed emission.

The Integral/IBIS Compton mode has proven to be a powerful tool to investigate polarization in astrophysical sources. Sensitive polarimetry in hard X rays/soft $\gamma$-ray with near-future instruments will provide invaluable means to explore particle acceleration to extreme energies around neutron stars and black holes.

\acknowledgements{This paper is based on observations with INTEGRAL, an ESA project with instruments and science data centre
funded by ESA member states (especially the PI countries: Denmark, France, Germany, Italy, Spain, and Switzerland), Czech Republic and Poland with participation of Russia and USA.}



\clearpage

\begin{deluxetable}{cc}
\tablewidth{0pt}
\tablecaption{Crab neutron star phase intervals\label{phases}}
\tablehead{\colhead{Name}& \colhead{Phase band}}
\startdata
$ \rm P_1$ & $0.88 < \phi < 0.14$ \\
$ \rm B  $ & $0.14 < \phi < 0.25$ \\
$ \rm P_2$ & $0.25 < \phi < 0.52$ \\
$ \rm OP $ & $0.52 < \phi < 0.88$ \\
\enddata
\end{deluxetable}

\begin{deluxetable}{cccc}
\tablewidth{0pt}
\tablecaption{Polarization angle and fraction with respect to the Crab pulsar phase $\phi$\label{results}}
\tablehead{\colhead{Crab pulsar phase interval}& \colhead{polarization angle}& \colhead{polarization fraction} & \colhead{chance probability}}
\startdata
 $\rm P_1~and~P_2$  &  $70^\circ \pm 20^\circ$     & $0.42 _{-0.16}^{+0.30}$    & $33.5 \%$ \\
 $\rm OP$           &  $120.6^\circ \pm 8.5^\circ$ & $> 0.72$ \tablenotemark{a} & $0.26 \%$ \\
 $\rm OP~and~B$     &  $122.0^\circ \pm 7.7^\circ$ & $> 0.88$ \tablenotemark{a} & $0.10 \%$ \\
 $\rm all$          &  $100^\circ \pm 11^\circ$    & $0.47 _{-0.13}^{+0.19}$    & $2.8 \%$  \\
\enddata
\tablenotetext{a} {The lower limits for the polarization fraction are given at the 95 \% confidence level.}
\end{deluxetable}

\clearpage

\begin{figure}
   \plotone{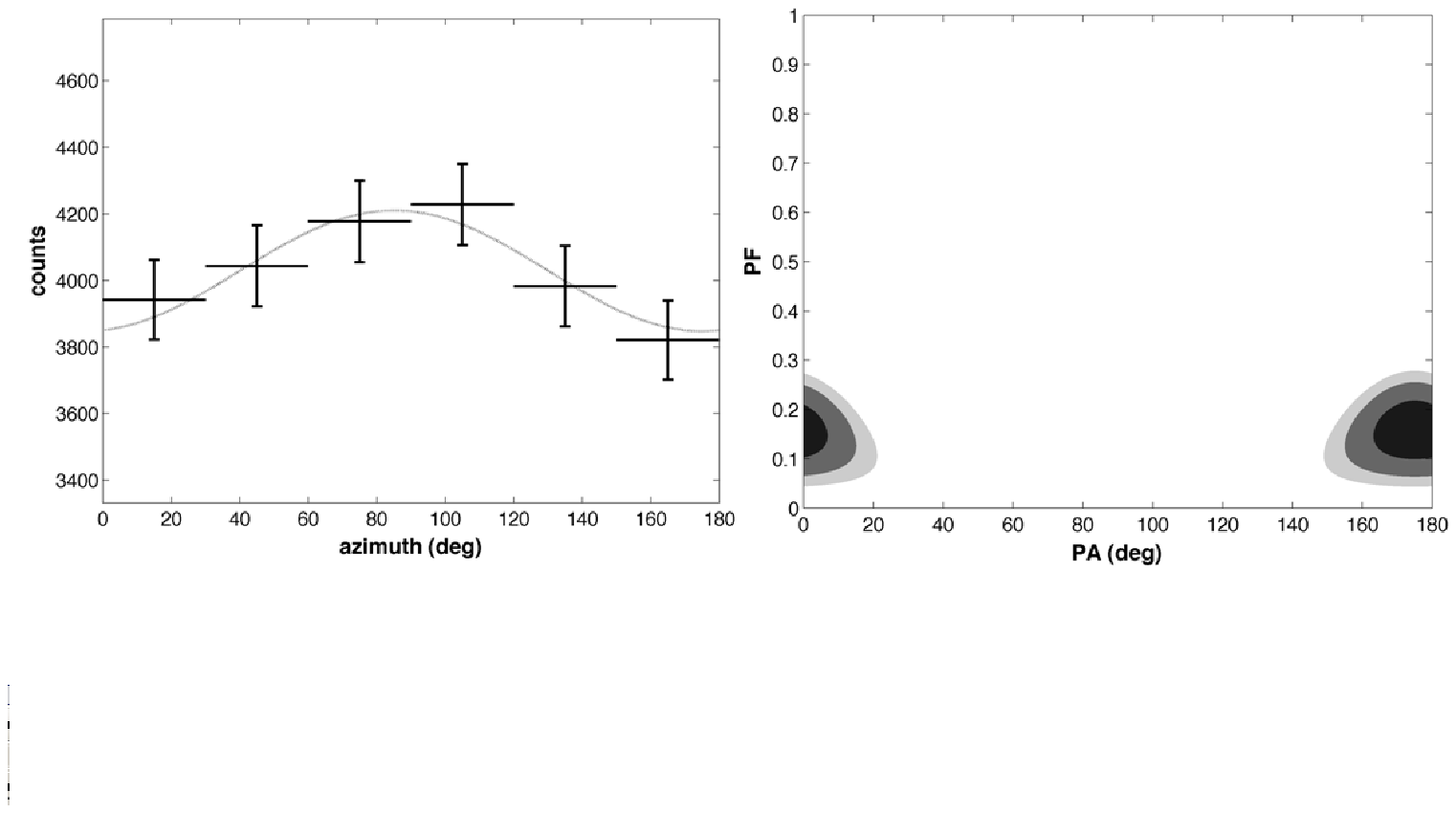}
   \caption{Azimuthal profile, modulation angle, PA, and fraction, PF = $a_0/a_{100}$, measured for a sample of unpolarized spurious events obtained during the Crab observations. The error bars for the profile are at one sigma. The 68\%, 95\%, and 99\% confidence regions are shaded from dark to light gray.}
   \label{spurious}
   \end{figure}

\end{document}